# Lumped-Element Model of THz HEB Mixer Based on Sputtered MgB$_2$ Thin Film

C. Yoo[1], C. Kim[1,a)], D. P. Cunnane[1], and B. S. Karasik[1,b)]

[1] *Jet Propulsion Laboratory, California Institute of Technology, Pasadena, CA 91109 USA*

We present a comprehensive analysis and experimental study of THz hot-electron bolometer (HEB) mixers made from 40-nm-thick sputtered magnesium diboride (MgB$_2$) thin films on high-resistivity silicon substrates. Using a lumped-element bolometric model, we achieve strong quantitative agreement with measurements of conversion gain, noise temperature, and local-oscillator (LO) coupling to the HEB devices. Our analysis shows that the sensitivity of current HEB devices is primarily limited by on-chip optical losses, with both Johnson and thermal-fluctuation noise contributing significantly to the overall noise temperature. Simulations of an optimized device with near-ideal optical coupling suggest that Johnson noise remains a substantial factor even with improved coupling. Further reduction of the noise temperature may require additional suppression of Johnson noise (via improved intrinsic conversion gain) beyond optimizing optical coupling efficiency. We emphasize the importance of accurate modeling to achieve good numerical agreement with experiments, thereby enabling understanding of the causes of sensitivity loss.

[a)] Current affiliation: HRL Laboratories, Malibu, CA 90265, USA. This research was performed while C. Kim was at the Jet Propulsion Laboratory, California Institute of Technology, Pasadena, CA 91109 USA.

[b)] Corresponding author. Electronic mail: boris.s.karasik@jpl.nasa.gov

## I. INTRODUCTION

Lowering the noise temperature of the THz superconducting hot-electron bolometer (HEB) mixer toward the quantum limit is crucial for enhancing astrophysical observations. Reducing the noise temperature by a factor of $N$ in a single HEB mixer effectively equals a $N^2$-fold increase in the scale of imaging arrays, which is considerably more expensive. So far, significant effort has focused on improving the optical design of the mixers to minimize external losses outside the HEB device. However, this alone does not ensure that the sensitivity reaches the quantum limit. Important experimental questions remain unanswered: how significant is the intrinsic (thermal) mixer noise relative to the quantum limit, and how can we systematically quantify and reduce it to achieve optimal performance?

Answering these questions has been challenging, primarily because developing practical, quantitatively precise models for HEB mixers remains difficult. Although HEB mixers have existed for thirty years, a persistent gap remains between mixer theory and the experimentally available HEB devices. The first HEB mixer [1], demonstrated on a large-area Nb device under an external magnetic field that induced a uniform resistive state, provided a solid foundation, but lacked practical significance due to its small intermediate frequency (IF) bandwidth of ~ 100 MHz. The HEB mixer theory described in [1] used a rudimentary bolometric model to predict the IF bandwidth successfully, but only a rough estimate of noise temperature and conversion gain based on a simplified model of electro-thermal feedback. A more sophisticated bolometric model with an advanced treatment of the self-heating effect was later developed for uniform HEB devices [2,3], but the field had shifted toward NbN-based HEBs [4-9] exhibited much shorter electron-phonon relaxation times and hence a practically valuable IF bandwidth up to ~ 2 GHz.

However, the high sheet resistance of thin NbN films (several hundred ohms per square) required sub-micron device lengths, leading to non-uniform resistive states caused by the formation and dynamics of resistive domains, whose size is determined by thermal diffusion and thus by the contact conditions (superconducting or normal). A similar inherent non-uniformity in sub-micron-long Nb [10-12] and Al diffusion-cooled HEB mixers, [13-15] leads to localized electron heating that greatly affects key mixer metrics, including noise performance and IF bandwidth, and thus presents a major challenge for modeling. For example, in "pure" diffusion-cooled HEBs, where cooling occurs predominantly via diffusion into contacts, the IF bandwidth $\Delta f_{IF}$ strongly depends on the device length $L$, following the relation $\Delta f_{IF} \sim 1/L^2$ [11,12,16], reflecting the characteristic diffusion time across the device. In NbN HEBs, however, the diffusion effect is less pronounced, and the IF bandwidth is mainly set by the combination of the electron-phonon relaxation time and the phonon escape time [7]. Nevertheless, diffusion remains a factor that influences the development of the resistive hot-spot and the boundary conditions at the contacts, affecting the current-voltage characteristics (IVC) and other mixer features.

To address this non-uniformity, the distributed electron-temperature or "hot-spot" model was introduced [17,18]. In this model, heterodyne mixing is linked to variations in the length of a localized hot spot where the electron temperature distribution is calculated by numerically solving coupled heat-balance equations under a specified local-oscillator (LO) power. The key mixer characteristics are then determined based on the resulting spatial distribution of the electron temperature. While the hot-spot model offers a general framework for understanding the mixing behavior of non-uniform HEB devices and has, in several cases [19-21], shown good alignment with the experimentally measured mixer characteristics of NbN HEB mixers, its practical use for routine analysis and device optimization remains limited due to the complexity of accurately modeling spatial profiles of electron temperature and extracting useful parameters for device optimization. Furthermore, defining the quantum limit is unclear within this framework because the model considers the conversion gain and noise temperature as spread along the device length, making it harder to distinguish between the optical loss caused by non-uniform absorption and the intrinsic quantum noise contribution [22-24].

In this work, instead of increasing the complexity of theory to account for non-uniformity in HEB mixers, we examine and analyze uniform HEB mixers using a simple lumped-element bolometric model [2,3]. Although the lumped-element bolometric model has successfully described the direct detection of transition-edge sensors (TESs), its quantitative applicability to heterodyne-mixer operation in HEBs remains unclear, primarily because uniform HEB devices are rare. Recently, we identified HEBs fabricated from sputtered magnesium diboride (MgB$_2$) films that exhibit exceptional uniformity [25], providing a unique platform for experimentally verifying the bolometric mixing model.

MgB$_2$ has emerged as a promising material for THz HEB mixers due to its high $T_c$ (~39 K) and short electron-phonon relaxation time, which enable operation at higher cryogenic temperatures and a wider IF bandwidth (~10 GHz), respectively. Following the first demonstration of the MgB$_2$ HEBs [26-28] based on films grown by molecular beam epitaxy (MBE), high temperature operation and wide IF bandwidth have been demonstrated in HEB devices [29-33] based on thin films produced by the Hybrid Physical-Chemical Vapor Deposition (HPCVD) [34,35]. Unlike HPCVD films, sputtered MgB$_2$ films [25] exhibit significantly better spatial uniformity due to a finer, polycrystalline structure with more uniform grain distribution and a smoother surface roughness (below 0.5 nm). The HEB devices made from these films show highly reproducible mixer characteristics and do not display the irregular features in the output noise that are often observed in HEB devices made from HPCVD-grown films. These irregular features are usually linked to non-uniformity caused by inter-grain boundary micro-switching events [29]. Furthermore, the low electron diffusivity value $D \leq 1$ cm$^2$/s in these sputtered films [36] and the relatively large length of our current devices ($L \approx 0.75$μm) give a diffusion relaxation time [16] $\tau_{diff.} = L^2/(\pi^2 D) \approx 600$ ps, much longer than the relaxation time observed in the experiment ($\approx 60$ ps), which also suggests that the lumped-element bolometric model [2,3] is applicable.

In the following sections, we present comprehensive modeling results using the lumped-element model along with a systematic analysis of various noise sources. This approach validates the use of a lumped-element bolometric mixing model, which supports its broader application for spatially non-uniform devices. Additionally, our model analysis enables us to directly estimate how close the mixer sensitivity is to the quantum limit, identify intrinsic noise mechanisms, and evaluate internal losses, providing useful guidance for further device improvement. As shown below, the model accurately reproduces both the bias dependence of the conversion gain and noise temperature measured from experimental IVCs of the HEB devices. Measurements at 2.5 THz confirm that these uniform MgB$_2$ HEBs perform in close agreement with bolometric model predictions.

## II. LUMPED-ELEMENT MODEL FOR HEB MIXERS

The lumped-element bolometric model [2,3] for uniform HEB mixers provides analytical expressions for their characteristics, assuming that the HEB mixer functions a lumped element that can be characterized by a single electron temperature $T_e$. In this framework, the device resistance $R(T_e)$ is a function of only the uniform $T_e$, and the heterodyne mixing is described solely in terms of variations in $T_e$ calculated from a relevant heat-balance equation (see Appendix A for an example).

An important factor when applying the lumped-element model to uniform HEB mixers is the condition under which the devices exhibit uniform heating. Strictly speaking, uniform heating can only occur in normal-metal regions. In the Nb HEB devices discussed earlier, the resistive state arises from vortex motion: moving vortices generate an electric field collinear with the transport current, resulting in an effective resistivity [37] smeared over the device area as the vortices cross the typical few-micron-long device at a high rate of several km/s [38]. At the microscopic level, a lattice of normal-metal vortex cores is embedded within a superconducting medium, so only a portion of the physical volume is occupied by the normal-metal vortex cores. The density of these cores increases with higher LO power and Joule heating as the HEB approaches the normal state. This effect was observed in an early work [39] where the effective thermal (electron-phonon) conductance was found to depend on the bias point. Also, a similar effect occurs in non-uniform NbN HEB, where it appears through the spatially non-uniform absorption of radiation with frequency $f < 2\Delta/h$, where $\Delta$ is the residual energy gap in the superconducting material outside the normal areas (vortex cores or resistive domains) [20]. The manifestation of this is the shape of the IVCs that cannot be simulated by a change in the physical temperature alone, as the application of radiation power does not always lead to a uniform increase of $T_e$ along the device. The effect of the non-uniform radiation absorption diminishes when the physical temperature of the HEB gets closer to $T_C$ [40] or when the radiation frequency gets higher [41]. When these two conditions are met, an appropriate increase in physical temperature can simulate the effect of LO pumping, and the HEB's noise properties are expected to follow the predictions of the lumped-element model closely.

Overall, the implication is that the accuracy of the lumped-element model may depend on both the proximity of $T_e$ to $T_C$ and the radiation frequency. In our analysis, we incorporate these effects by introducing a bias-dependent THz absorption efficiency, which will be explained in more detail in the following sections. Below, we first provide an overview of the lumped-element mixer theory. The detailed derivation of the equations for mixer characteristics was introduced in earlier studies of the lumped-element model [2,3], but is presented here for reference in the present model analysis.

First, the expression for the mixer conversion gain $\eta_M$ is given by:

$$\eta_M(\omega) = 2\alpha C^2 \frac{RR_L}{(R+R_L)^2} \frac{P_{LO}}{P_{DC}} \frac{1}{\left(1+C\frac{R-R_L}{R+R_L}\right)^2} \frac{1}{1+(\omega\tau^*)^2}. \tag{1}$$

Here, $\alpha$ is the RF coupling factor of THz radiation to the mixer, $C$ is the self-heating parameter, $R$ is the device ohmic resistance, $R_L = 50\ \Omega$ is the IF load impedance, $P_{LO}$ is the THz LO power coupled to the HEB device, $P_{DC}$ is the dissipated Joule power, and $\tau^*$ is the thermal time constant including the electrothermal feedback (ETF) effect (see Appendix B) given by [3]:

$$\tau^* = \frac{\tau}{1+C\frac{R-R_L}{R+R_L}}, \tag{2}$$

where $\tau$ is the thermal time constant of the bolometer in the absence of the ETF effect. The expression for the conversion gain in Eq. (1) is identical to that derived in [42] except for the definition of the dimensionless self-heating parameter $C = I^2 \frac{dR}{dT}/G$, where $I$ is the bias current and $G$ is the effective thermal conductance between electrons and the external environment.

The RF coupling factor $\alpha$ is determined by the impedance matching between the THz antenna impedance $Z_A$ and the device's RF impedance at the operating frequency. We adopt $Z_A \approx 80\ \Omega$ for our planar log-spiral antenna, based on numerical simulations of a similar geometry [43]. The conventional assumption for $Z_{RF}$ at frequencies well above the superconducting gap ($f \gg 2\Delta/h$) is that THz radiation is uniformly absorbed, giving $Z_{RF} = R_N$ where $R_N$ is the normal resistance. At frequencies below the gap, however, photons cannot break Cooper pairs and are instead absorbed by thermally excited quasi-particles whose density depends strongly on $T_e$ [20]. In this low-frequency regime, the RF impedance can be significantly smaller than $R_N$, as widely observed in HEB mixers [5,20,41,44]. This behavior is further complicated in MgB$_2$ due to the existence of two superconducting gaps, making it difficult to evaluate $Z_{RF}$ *a priori* at our operating frequency of 2.52 THz. Although more general theoretical treatments for $Z_{RF}$ exist [45], their implementation is not practical for our modeling purposes. Therefore, we adopt a simplified empirical approach by taking $Z_{RF} = \beta R$, where $R$ is the Ohmic device resistance and $\beta < 1$ is a constant that captures the reduced THz absorption and its bias dependence. This bias-dependent device impedance effectively models the varying degree of uniform heating along the DC bias. It reproduces the observed mixer performance with excellent agreement (see Sec. IV). Lastly, to account for any additional *bias-independent* optical losses on the device chip, we introduce another scaling parameter $\alpha_0$ to determine the overall RF coupling factor $\alpha = \alpha_0 \cdot 4Z_{RF}Z_A/(Z_{RF} + Z_A)^2$.

The self-heating parameter $C$ is a valuable indicator of the sensitivity of a lumped-element HEB device and can be extracted from the IVC of the device:

$$C = \frac{Z-R}{Z+R}, \qquad (3)$$

where $Z = dV/dI$ is the differential resistance at DC. For example, Fig. 1 illustrates the dependence of $C$ on the position of the bias point of a typical HEB device at different levels of LO pumping power. When $Z > 0$ ("optimally" pumped and over-pumped cases shown in Fig. 1(a)), the self-heating parameter is always less than 1 ($C < 1$), according to Eq. (4) (see the green and red curves in Fig. 1(b)). As $Z$ approaches infinity, $C$ is approximately equal to 1 (yellow curve for the under-pumped case). When $Z < 0$, $C$ can significantly exceed 1 (blue curve for the minimally pumped case); a similar IVC is widely observed in the Transition-Edge Sensors (TES) with negative electrothermal feedback [46]. For TES devices, because of the low impedance of the read-out amplifier (SQUID) and the small signal bandwidth of $10^3$-$10^5$ Hz, it is relatively easy to stabilize the bias point when $Z < 0$ by ensuring $R_L < |Z|$. However, the latter condition is difficult for HEB mixers, as a negative $Z$ usually causes parasitic oscillations, smearing the bias point, and degrading mixer performance. Thus, $C$ is always less than 1 in a typical operation, and the electrothermal feedback is always positive [47].

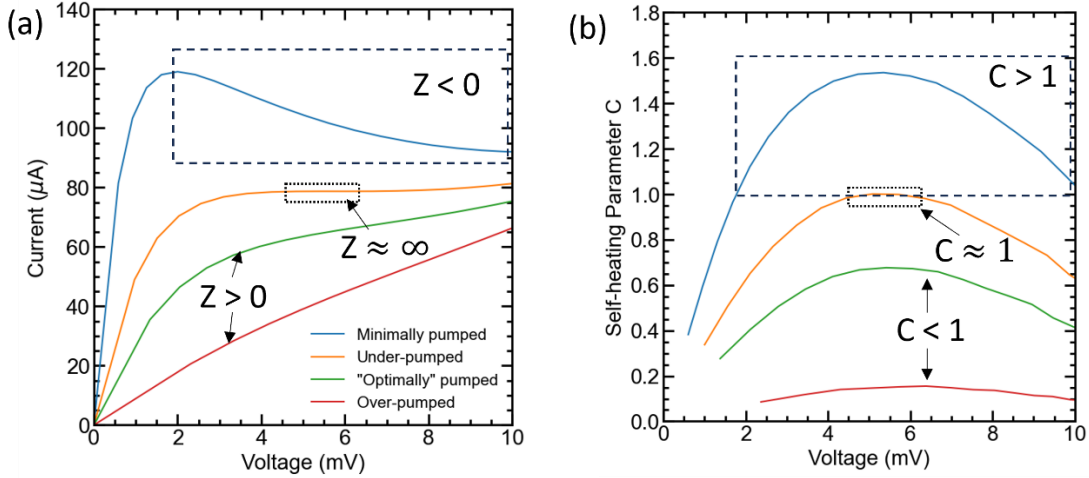

FIG. 1. (a) Simulated current-voltage characteristics (IVC) of an HEB device at different levels of LO pumping. (b) Self-heating parameter extracted from the simulated IVCs.

The treatment of the mixer noise is based on a model [2,3] that takes into account both Thermal Energy Fluctuations (TEF) noise (aka "phonon noise") and Johnson noise. The output noise powers for the TEF noise ($P_{IF}^{TEF}$) and the Johnson noise ($P_{IF}^{J}$) are given by the following equations:

$$P_{IF}^{TEF}(\omega) = \frac{2k_B T_e^2 GB}{P_{DC}} \frac{RR_L}{(R+R_L)^2} \frac{C^2}{\left(1+C\frac{R-R_L}{R+R_L}\right)^2} \frac{1}{1+(\omega\tau^*)^2} \quad (4)$$

and

$$P_{IF}^{J}(\omega) = k_B T_e B \frac{4RR_L}{(R+R_L)^2} \frac{1}{\left(1+C\frac{R-R_L}{R+R_L}\right)^2} \frac{1+(\omega\tau)^2}{1+(\omega\tau^*)^2}, \quad (5)$$

where $B$ is the bandwidth of the IF signal.

Expressing the output noise power in temperature units ($T_{IF}^{TEF(J)} = P_{IF}^{TEF(J)}/(k_B B)$) and dividing it by $2\eta_M$ given by Eq. (1) yields the expression for DSB mixer noise temperature for each contribution:

$$T_M^{TEF} = \frac{T_e^2 G}{2\alpha P_{LO}} \quad (6)$$

$$T_M^{J}(\omega) = \frac{T_e P_{DC}}{\alpha C^2 P_{LO}}[1+(\omega\tau)^2]. \quad (7)$$

The total DSB mixer noise temperature is then $T_M^{DSB} = T_M^{TEF} + T_M^{J}$. In practice, the sensitivity of HEB mixers is often quantified in terms of the receiver noise temperature:

$$T_{REC}^{DSB} \approx \frac{T_M^{DSB}}{\eta_{opt}} + \frac{T_{IF}}{2\eta_{opt}\eta_M} \quad (8)$$

where the first term represents the mixer noise contribution with the optical coupling efficiency $\eta_{opt}$ of the receiver outside the HEB device, and the second term represents the noise contribution $T_{IF}$ from the IF chain.

The inclusion of quantum noise [22-24] can be crucial for evaluating the overall sensitivity limit of HEB mixers when the classical (thermal) contributions given in Eqs. (6) and (7) approach their theoretical limit. The expression for the classical DSB limit $T_M^{CL}$ was derived [3] for the operating condition when $P_{LO} \gg P_{DC}$ and $C \cong 1$ that reduces the Johnson noise contribution given in Eq. (7) to zero:

$$T_M^{CL} = nT_C/2, \qquad (9)$$

where $n$ is the exponent in the temperature dependence of the thermal flux between electrons and phonons (or the substrate). The value of $n$ is 4 in ultrathin Nb film [48,49], ranges from 3.6 [4] to 5.8 [50] in disordered NbN films with $T_C \approx 10$ K, and is likely to be $n \approx 3$ for MgB$_2$ films [30,51] (see Appendix A). Thus, for NbN HEBs with $T_C \approx 10$ K, $T_M^{CL}$ should be less than 30 K. Near this limit, the noise temperature limit of HEB mixers must include the quantum noise contribution given by (in terms of DSB value):

$$T_M^{QL} = hf/2k_B, \qquad (10)$$

where $h$ is the Planck constant, $f$ is the radiation frequency, and $k_B$ is the Boltzmann constant [52-54]. Over the 1 to 5 THz range, this variation ranges from 24 to 120 K, resulting in an overall limit $T_M^{QL} + T_M^{CL}$ of 54 to 150 K for NbN HEBs in the same frequency range. However, the experimental mixer noise temperature of the most sensitive NbN HEB mixers currently available [9] is at least 3-5 times higher than this fundamental limit. For our model analysis at 2.52 THz, the quantum contribution ($T_M^{QL} \approx 61$ K) is negligibly small relative to the measured DSB receiver noise temperature (3,000 – 3,500 K) and therefore is safely ignored. Lastly, it should be noted that the expression for the classical limit given in Eq. (9) was obtained under the restrictive conditions with $P_{LO} \gg P_J$ and $C \cong 1$, which can be challenging to realize experimentally. Indeed, as we show below, this inequality does not hold in our MgB$_2$ HEB devices either, making the classical contribution of thermal noise much larger than what Eq. (9) predicts.

### III. DEVICE AND EXPERIMENTAL SETUP

#### A. Film Growth and Device Fabrication

The device fabrication started with growing of a 40-nm MgB$_2$ film on a high-resistivity silicon (Si) substrate [25]. First, the Si substrate was pre-coated with a 30-nm SiN$_x$ buffer layer (Fig. 2(a)) to prevent the formation of Mg$_2$Si grains during post-deposition annealing. After the pre-coating, the magnesium (Mg) – boron (B) composite film was co-deposited using magnetron sputtering at room temperature, with a small RF bias applied to the substrate to ensure surface smoothness of the composite film. Next, a 30-nm boron cap layer was grown to prevent Mg evaporation and oxidation during annealing. Finally, the film was annealed at approximately 600 °C in nitrogen gas for 2 to 10 minutes. After annealing, the 40-nm MgB$_2$ film exhibited a resistivity of ~ 85 µΩ·cm, a critical current density of ~10 MA/cm$^2$ (at 4.2 K), and a critical temperature of ~32 K.

The HEB devices were fabricated following the steps detailed in [29,30]. The process began with defining the contact pads for the HEB bridge. First, the boron cap layer was removed by dry etching, and a Ti/Au/Ti layer was deposited across the entire wafer using e-beam evaporation. Next, the contact pads were patterned on the Ti/Au/Ti layer using photolithography, and the metal layer outside the contact pattern was removed with dry etching. The exposed $MgB_2$ was then etched down to $d \sim 7$ nm using Ar ion milling [29,30]. Immediately after milling, a $SiN_x$ layer was deposited to protect the newly exposed $MgB_2$ layer. Next, the HEB bridges were defined using photolithography. Another round of dry etch and ion milling was performed to remove the $SiN_x$ layer and the remaining $MgB_2$ layer outside the HEB bridges. After milling, the exposed sidewalls of the HEB bridges were passivated with an additional $SiN_x$ layer. Finally, a spiral antenna pattern was defined via photolithography and lift-off (Fig. 2(b)). To further protect the $MgB_2$ in the HEB bridge area, an additional $SiN_x$ layer was deposited on the 5 μm × 5 μm window at the center of the spiral antenna.

We investigated three HEB devices with a nominal bridge width of 2 μm and a length of 0.75 μm (Table 1). While fabrication errors resulted in a significant spread in the normal resistances of the devices (Fig. 2(c)), all three HEB devices exhibited qualitatively similar $R(T_e)$ dependence with $T_C \sim 22\text{-}23$ K (Fig. 2(c)). The significantly lower $T_C$ of the fabricated devices compared to $T_C \approx 32$ K for the bulk film is attributed to degradation during ion milling of the thinned-down HEB bridges. The $R(T_e)$ curves of all three devices also exhibit second transitions near $\approx 27$ K attributed to the transitions in the contact pads that still host the 40-nm-thick $MgB_2$ after the fabrication. The presence of the $MgB_2$ material with a higher $T_C$ than

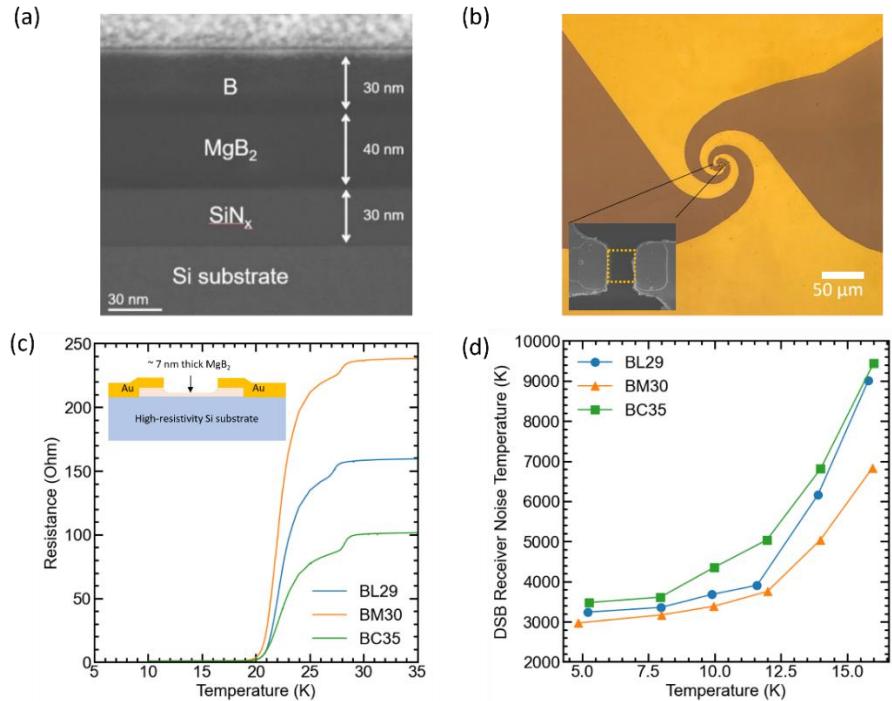

FIG. 2. (a) Transmission Electron Microscope (TEM) image of the 40-nm $MgB_2$ film used for the fabrication of the HEB devices. (b) Optical microscope image showing the spiral antenna of the HEB device, with the inset showing the HEB bridge. (c) $R(T)$ curves for the three HEB devices tested in this study, with the inset showing the schematic for the vertical structure of the device (not to scale). (d) Temperature dependence of the uncorrected DSB receiver noise temperature.

8© 2026 California Institute of Technology. Government sponsorship acknowledged.

in the bridge suggests that the Andreev reflection [55] is likely to be present, suppressing the out-diffusion of the electrons. Based on the nominal dimension of the HEB bridge and the electrical properties measured for BL29, a resistivity $\rho_N$ of ≈ 300 μΩ·cm (corresponding to a sheet resistance $R_s$ ≈ 400 Ω/sq) and a critical current density $j_c$ of ≈ 2 MA/cm$^2$ were estimated. The estimated critical current density is several times lower than that of the unprocessed 40-nm MgB$_2$ film, owing to degradation of film quality during fabrication. Note, although $j_C$ > 10 MA/cm$^2$ is required for achieving good sensitivity in the HEB mixer made from the HPCVD-grown MgB$_2$ film [56], this does not indicate the inferiority of the sputtered films for making an HEB mixer from the current films; the critical parameter $j_C^2 \rho_N$ determines the sensitivity, and the value $j_C^2 \rho_N$ is comparable in the HPCVD-grown and sputtered films, with the significant $j_C$ value in the HPCVD-grown films [57] compensated by a smaller $\rho_N$ ~ 5-10 μΩ·cm value. Regarding mixer characteristics, all three devices exhibit DSB receiver noise temperatures ranging from 3,000 K to 3,500 K at 5 K, with similar temperature dependence (Fig. 2(d)). They have IF bandwidths of 2-3 GHz (see Appendix B for the detailed measurements for BL29) and coupled LO powers of 1-2 μW, as determined from the isothermal analysis [39,40].

TABLE I. Physical properties of tested HEB devices.

| Device | Nominal Thickness $d$(nm) | Nominal $W$(μm) × $L$(μm) | $T_C$ (K) | $\Delta T_C$ (K) | $R_N$ (Ω) | $I_C$ (μA) |
|---|---|---|---|---|---|---|
| BL29 | 7 | 2 × 0.75 | 22.6 | 5.2 | 160 | 330 |
| BM30 | 7 | 2 × 0.75 | 22.3 | 4.4 | 240 | 330 |
| BC35 | 7 | 2 × 0.75 | 22.8 | 7.1 | 100 | 240 |

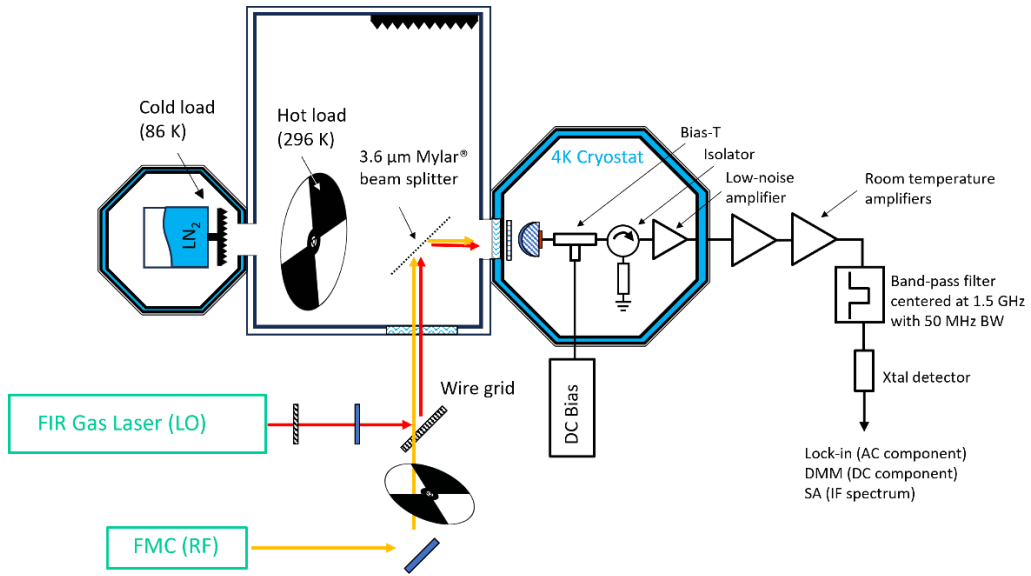

FIG. 3. Schematic of the dual lock-in setup for the bias dependence measurement.

## B. Setup for Noise Temperature and Bias Dependence Measurements

The noise temperature of the receiver and its temperature dependence, shown in Fig. 2(d) were characterized using the Y-factor method with a vacuum setup similar to the one in Fig. 3. For this measurement, the second RF source from the frequency multiplier chain (FMC) was not used and can be ignored. First, the mixer device was bonded to a high-resistivity silicon (Si) lens coated with a Parylene anti-reflection layer, and the device-lens assembly was mounted on a copper mixer block installed on the 4-K plate of a liquid-helium cryostat. The device was pumped with a linearly polarized far-infrared (FIR) gas laser LO at 2.522 THz (red arrow). The LO power was adjusted for optimal pumping using a rotating wire grid that acts as an attenuator and is quasi-optically coupled to the device via collimating optics and a 3.6-µm thick Mylar beamsplitter. For the Y-factor, thermal radiation from 295 K (hot load) and 86 K (cold load) black bodies was coupled to the device through the beamsplitter and a 0.5-mm thick HDPE window of the dewar. At the dewar window (at the 77-K stage), a thermal IR filter (200-µm thick Zitex™ 108) was also installed to block the IR background. To suppress potential direct detection effects [58-60] prevalent for small-volume HEB mixers with low LO power requirements, a THz bandpass filter made from a thin freestanding copper mesh film with a center frequency near 2.5 THz (with a transmission of ~ 72 % at 2.522 THz and a bandwidth of ~ 10 %) was placed in front of the mixer block (at the 4-K stage). Since the measurements, an improved version of the metal-mesh film offering better transmission (> 90 %) at 2.522 THz, has become available [61], but was not used in the presented experimental results.

Inside the dewar, a bias-T and a cryogenic, high-gain, low-noise microwave amplifier (LNA) were installed at the 4-K stage for DC biasing of the HEB and low-noise amplification of the IF power, respectively. An L-band isolator was also placed between the mixer and the LNA input port to prevent oscillations in the IF circuit caused by potential impedance mismatch. The noise temperature of the entire IF chain was measured to be $T_{IF} \approx 2K$ at 4.2 K. Outside the dewar, the IF signal was further amplified by a chain of room-temperature microwave amplifiers and filtered with a microwave bandpass filter centered at 1.5 GHz with a 50-MHz bandwidth. For the noise temperature measurement shown in Fig. 2(d), a calibrated power meter was used to measure the IF power as the hot/cold loads were manually switched using an optical chopper to extract the Y-factor at the optimal LO pumping and DC bias points.

The bias dependence of the receiver noise temperature and conversion gain was measured for BL29 using the exact setup shown in Fig. 3. In this measurement, a small RF signal was generated by the FMC near 2.522 THz for the simultaneous monitoring of the mixing signal at the IF frequency of 1.5 GHz. The RF signal was coupled through the wire grid and the Mylar beamsplitter, which significantly attenuated the RF power coupled to the device to below 1 % of the required LO power (~ 1 – 2 µW). For the bias-dependence measurements, the HEB device was first pumped to the optimal LO power level at which the maximum Y-factor (at the optimal DC bias points) was observed in the previous noise temperature measurements. At this optimal LO pumping level, the DC bias was scanned, and the device's current and voltage, Y-factor, and IF signal were simultaneously recorded at each bias point using digital multimeters (DMMs) and two lock-in amplifiers. To enable faster readout of the lock-in amplifiers, the calibrated power meter at the output of the microwave bandpass filter was replaced with a fast crystal diode. For the Y-factor response, both the DC and AC components of the crystal detector were measured using the DMM and the first lock-in amplifier referenced to the optical chopper for the hot/cold load, respectively. The measured responses were calibrated using the Y-factor previously determined from the actual DC responses obtained by slow, manual switching of the hot/cold loads. For the IF response, the second lock-in amplifier was referenced to the second optical chopper for the RF signal at a chopping frequency different from that of the hot/cold load chopper. Throughout the scan, the IF spectra were monitored with a spectrum analyzer to ensure the absence of intermodulation components resulting from parasitic oscillations.

## IV. MEASUREMENT AND MODEL RESULTS

### A. Measurement Results

Figure 4 shows the measured bias dependencies of the receiver noise temperature and conversion gain. The receiver noise temperature was extracted from the Y-factor using brightness temperatures $T_{HOT}$ and $T_{COLD}$ derived from the Callen-Welton relationship for the hot (295 K) and the cold (86 K) black-body loads [52]. At each bias point, the Y-factor was calculated from the measured output noise powers in response to hot/cold load radiation. ($Y = P_{out}^{HOT}/P_{out}^{COLD}$), and the receiver noise temperature was then found as

$$T_{REC}^{DSB} = \frac{T_{HOT} - Y \cdot T_{COLD}}{Y-1}. \tag{11}$$

The bias dependence in Fig. 4(a) shows that the noise temperature monotonically decreases with bias and approaches a minimum of ≈ 3,200 K above a DC bias of ≈ 4 mV.

The receiver conversion gain was extracted using the U-factor method:

$$\eta_{REC} = \eta_{opt}\eta_M = \frac{U(T_{50\Omega} + T_{IF})}{2(T_{REC}^{DSB} + T_{HOT})}, \tag{12}$$

where $\eta_{opt}$ is the optical loss in the receiver, $\eta_M$ is the mixer conversion gain, $U$ is the U-factor [8,62] measured as the ratio between the output noise power and the reference output noise power when the mixer is at zero-bias, and $T_{50\Omega} + T_{IF}$ is the calibrated output noise temperature when the mixer is at zero-bias ($T_{50\Omega} \approx 5$ K at 5 K). An optical coupling loss $\eta_{opt} = -3.22$ dB was estimated based on the measured and computed optical losses of the receiver components, including the vacuum window, heat filter, THz bandpass filter, and Mylar beamsplitter. Figure 4(b) shows that the conversion-gain values calibrated using the U-factor method of Eq. (12) (red triangles) agree very well with the directly measured amplitudes of the 1.5 GHz IF tones in independent measurements (orange squares).

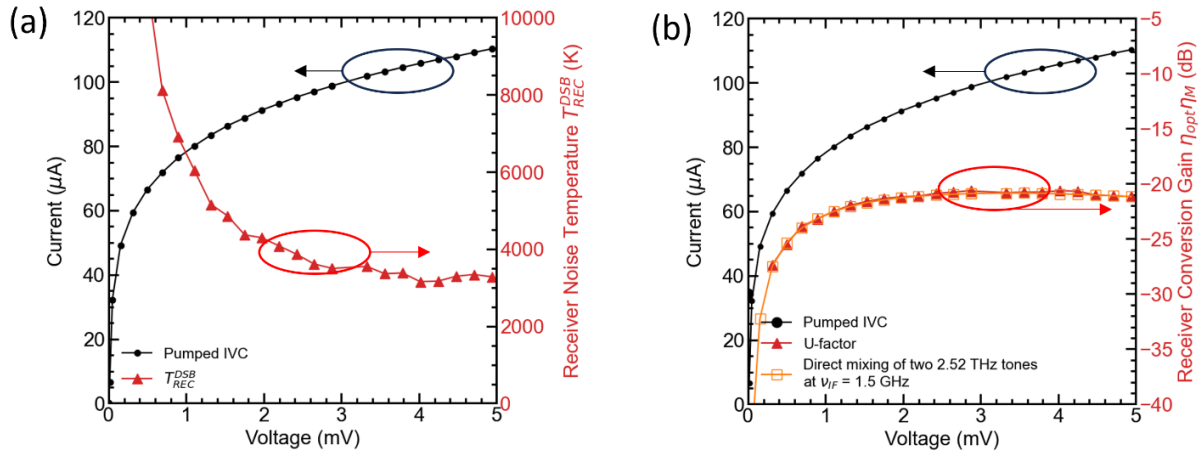

FIG. 4. (a) Experimental IVC of BL29 at optimal LO pumping at 5 K, along with the measured DSB receiver noise temperature at 5 K at 2.52 THz. (b) Experimental bias dependence of DSB receiver conversion gain including the optical loss in the front-end of the receiver.

### B. Model Results

Our model analysis began by extracting a smooth, analytic representation of the experimentally measured IVC. Theoretically, modeling HEB mixers can be approached from first principles by directly reproducing the mixer's IVC, as shown in [63]. In this approach, the electron temperature $T_e$ is initially determined across a range of $P_{DC}$ values at a given $P_{LO}$ by solving the relevant heat-balance equations (see Appendix A for the heat-balance equation used in our case). The device resistance $R$ at each $P_{DC}$ is then determined based on the experimental $R(T_e)$ dependence, which is typically modeled using an analytic expression shaped like a Fermi function to match the measured transition slope ($dR/dT_e$) at the critical temperature $T_c$ (see the dotted red line in Fig. 5(a) as an example). Subsequently, the current and voltage over the selected $P_{DC}$ range are calculated to generate the IVC for the given $P_{LO}$, from which all the mixer characteristics are derived.

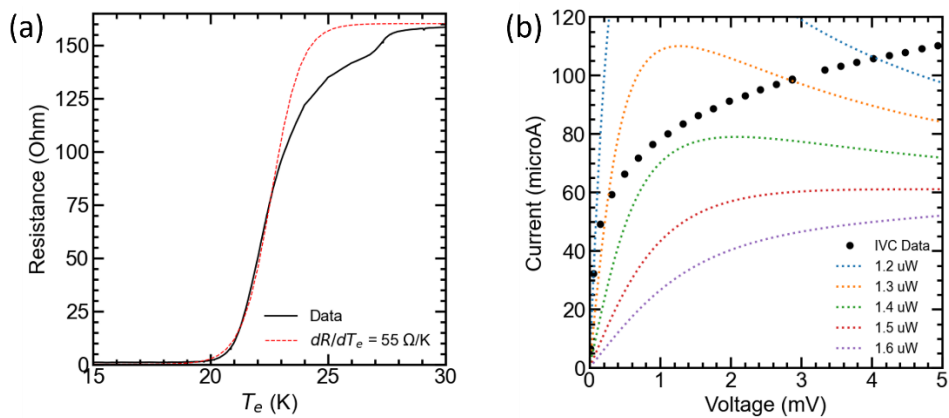

FIG. 5. (a) Experimental $R(T_e)$ curves and analytical functions used for IVC simulation for BL29. (b) Simulated IVCs with analytical $R(T_e)$ curve with $dR/dT_e$ = 55 Ohm/K at $T_e=T_C$ for various $P_{LO}$.

However, we found that this first-principles approach did not accurately reproduce the experimental IVC of our current HEB devices (Fig. 5). Specifically, using a single $R(T_e)$ curve that matches the experimental transition for $T_e < T_C$ (see the dotted red line in Fig. 5(a)) across relevant $P_{LO}$ ranges produced model IVCs that deviate significantly from the experimental data (Fig. 5(b)). We primarily attribute this discrepancy to bias-dependent absorption of THz radiation, which leads to varying degrees of uniform heating across the DC bias range at a fixed incident LO power, as discussed in the model introduction. Additionally, another possible explanation is the non-thermal current-bias dependence of the resistance $R = R(T_e, I)$, which is independent of Joule heating. Such behavior in HEBs was first reported in [39], but has been largely overlooked in subsequent lumped-element mixer modeling studies. A similar discrepancy has been observed in distributed model analysis for non-uniform NbN HEB mixers. [64]. It has been suggested that this discrepancy may arise from vortex-antivortex unbinding, as described by the Berezinskii-Kosterlitz-Thouless (BKT) theory [65]. In particular, the increase in resistivity of ultrathin NbN films has been attributed to the emergence of free vortices under current bias, leading to a nonlinear dependence of resistance on current. A successful agreement between modeled and experimental IVCs has been achieved within the framework of the BKT theory by incorporating the experimentally measured current-bias dependence of $R(T_e)$ [19,64]. For our modeling work, we instead begin with the experimental IVC for our analysis, thus avoiding reliance on an assumed $R(T_e)$ form.

For the numerical modeling, the experimental IVC was fitted with the following polynomial function: $I = aV^b + kV$ where $a$, $b$ (<0), and $k$ are fitting parameters. This polynomial fit accurately reproduced the experimental IVC (dashed black line in Fig. 6(a)). It minimized numerical artifacts from differentiating discretely sampled data, thereby enabling reliable extraction of both linear and differential resistances (solid and dashed red lines in Fig. 6(a)). Using these extracted resistances, the self-heating parameter $C$ was computed using Eq. (3) and is plotted as the solid red curve in Fig. 6(b). As the bias increases, $C$ quickly approaches ≈ 0.65 and remains relatively constant for bias voltages above ≈ 0.5 mV.

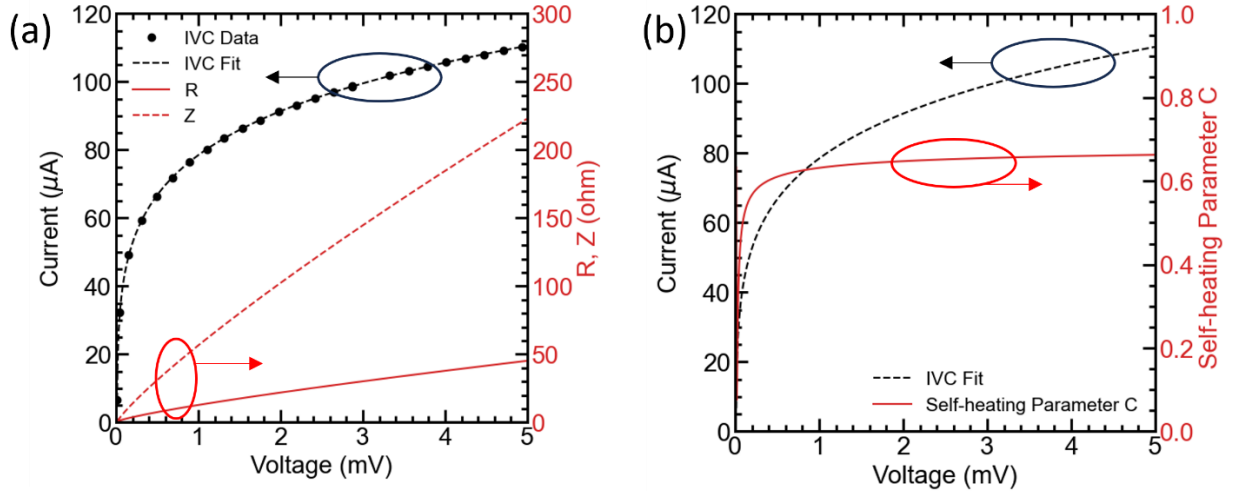

FIG. 6. (a) Model fit for IVC (black dashed line) and extracted linear resistance (solid red line) and differential resistance (dashed red line). (b) Modeled IVC (black dashed line) and extracted self-heating parameter (red solid line).

Subsequently, the bias dependence of the receiver conversion gain was modeled at $f$ = 1.5 GHz using Eq. (1) (Fig. 7(a)). For the RF coupling factor $\alpha$, we assumed $\alpha_0$ =0.21 and $\beta = 0.35$, which resulted in good agreement between the modeled conversion gain and the experimental data (Fig. 7(a)). However, these values yielded a relatively low estimated RF coupling factor of $\alpha \approx$ - 9.6 dB near the optimal operating point (> 3 mV). Therefore, a relatively large *incident* LO power $P_{inc}$ = 10.5 µW was used in the model to match the experimental data points for the *coupled* LO power $P_{LO} = \alpha P_{inc} \approx 1$ µW near these optimal bias points. The full bias dependence of $P_{LO}$ was calculated from the estimated $\alpha$ and given $P_{inc}$. The resulting curve for $P_{LO}$ (dashed black line in Fig. 7(b)) agreed very well with the experimental data points (blue stars in Fig. 7(b)) extracted using the isothermal method. The bias dependence of the mixer time constant $\tau^*$ was derived from experimental measurements (see Appendix C). Finally, a receiver optical loss of $\eta_{opt}$ = -3.22 dB was added to the modeled mixer conversion gain for comparison with the measured receiver conversion gain. As shown in Fig. 7(a), the modeled receiver conversion gain (solid red line) closely matches the experimental data (red triangles).

To model the receiver noise temperature, the mixer noise temperature was first calculated by adding the TEF and Johnson noise contributions, as given by Eqs. (6) and (7). The total receiver noise temperature was then determined using Eq. (10), which accounts for the receiver's optical loss of $\eta_{opt}$ = -3.22 dB and an IF-chain noise contribution of $T_{IF} \approx 2$ K. The electron temperature $T_e$ required for these calculations was estimated from the measured $R(T_e)$ curve using the ohmic resistance extracted from the IVC (Fig. 6(a)). The effective thermal conductance was assumed to be constant at $G$ = 200 nW/K, based on a constant-resistance-line analysis of IVCs measured at multiple bath temperatures (Appendix A). Although the model suggests $G$ may depend on $T_e$ and thus on the bias point, we use a constant value since $T_e$ is expected to vary only slightly (20-22 K) within the bias range.

Figure 7(c) shows that the modeled bias dependence of the receiver noise temperature (solid red line) matches the measured data (red triangles) well. The model captures the decrease in total noise temperature as bias increases, mainly due to the reduction in the TEF noise contribution (red dashed). The Johnson noise (red dotted) and the IF-chain noise (red dash-dot) stay roughly constant over most of the bias range, except at very low bias (< 0.5 mV), where the receiver conversion gain drops sharply (red curve in Fig. 7(a)). Near the start (~ 2.5 mV) of the noise plateau, the TEF and Johnson noise contributions are approximately equal. At the optimal operating points (> 3 mV), the Johnson-noise contribution (~1,500 K) begins to dominate, with the TEF contribution decreasing and the IF-chain noise contribution remaining relatively small (~250 K).

Our model analysis indicates that the mixer performance of our sputtered MgB$_2$ HEB devices is limited by significant optical loss, with an estimated optical coupling factor $\alpha = -9.5$ dB near the optimal operating points > 3 mV. Based on this estimate, only ~ 11 % of the incident THz power is expected to be absorbed by the HEB device. Since the optical losses of the components before the mixer, such as the cryostat window, heat filters, and beamsplitter, were experimentally characterized and included in the receiver optical loss ($\eta_{opt} = -3.22$ dB), we conclude that this substantial attenuation occurs within the device chips. In our model, this on-chip loss is characterized by low values of both the bias-independent coupling parameter $\alpha_0 = 0.21$ and the bias-dependent factor $\beta = 0.35$. Although identifying the precise origins of these losses is complex, the low value $\alpha_0$ indicates that significant losses occur outside the HEB bridge, regardless of the bias condition. The most probable causes are resistive losses in the antenna structure and, more importantly, poor contact at the interface between the antenna leads and the MgB$_2$ film. This interface is especially vulnerable in our fabrication process, where removing the protective boron cap layer before contact definition exposes the MgB$_2$ surface to possible damage and residual resistance. Similar contact-related sensitivity degradation has been observed in NbN HEB mixers [9,66,67], too, suggesting that improving the contact interface quality is likely the most direct approach to enhancing the mixer performance of our current HEB devices.

In addition to the extrinsic resistive loss captured by $\alpha_0$, the low value of $\beta$ indicates more fundamental inefficiencies in

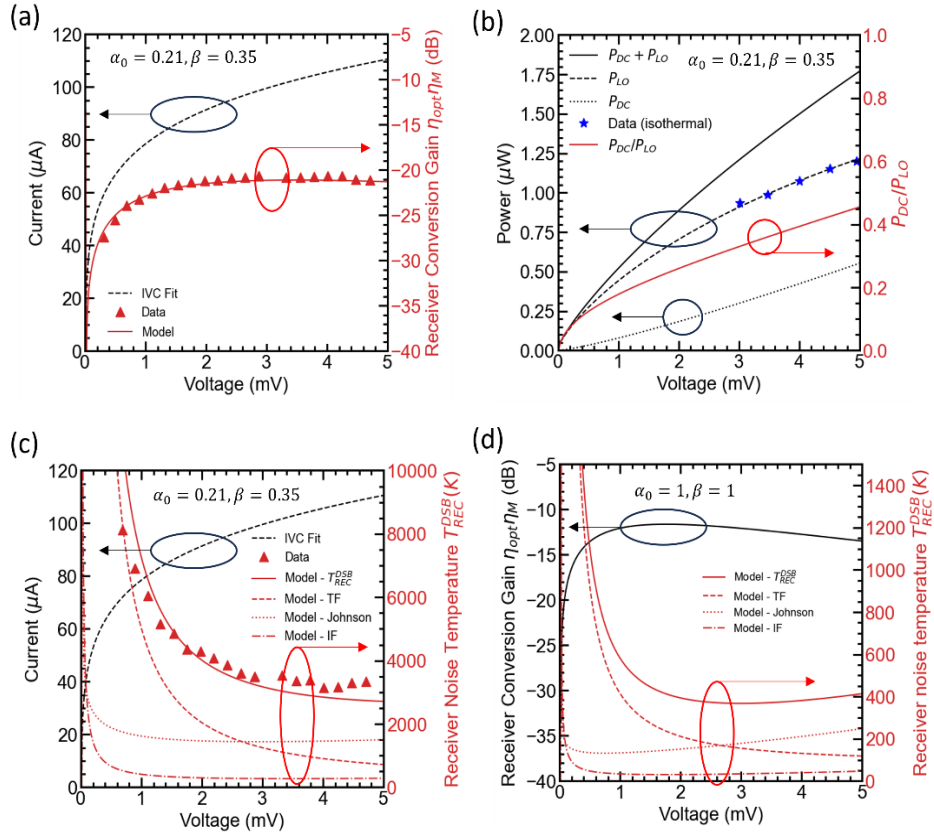

FIG. 7. Lumped-element model analysis results (a) Modeled IVC (black dashed line) and conversion gain (red solid line) with experimentally measured data points (red triangle). (b) Modeled total power (solid line) absorbed in the HEB device showing contributions from the LO power (dashed line) and dissipated Joule power (dotted line), compared with data points for $P_{LO}$ experimentally extracted using isothermal line analysis (blue stars). (c) Modeled receiver noise temperature (solid red line) showing contributions from thermal fluctuation noise (red dashed line), Johnson noise (red dotted line), and IF chain noise (red dashed-dotted line). (d) Model prediction for conversion gain (black solid line) and overall noise temperature (red solid line) of an optimized device with $\alpha_0 = 1$ and $\beta = 1$ with an incident LO power of 1 $\mu W$.

uniform heating within the HEB devices. In our model, the parameter $\beta$ was used to phenomenologically describe the device RF impedance that depends on the bias ($Z_{RF} = \beta R$). We adopted this relationship based on the hypothesis that uniform heating can occur to varying degrees depending on the density of the normal-metal cores within the superconducting medium, which is influenced by both electron temperature and radiation frequency. From this modeling perspective, the low value of $\beta$ leads
15

to a reduced device RF impedance, thereby significantly contributing to the overall optical loss via impedance mismatch between the integrated THz antenna and the HEB bridge. Physically, $\beta$ roughly reflects the intrinsic electromagnetic response of the superconducting material to high-frequency radiation, which can manifest as a frequency dependence of the optical coupling efficiency (and hence in noise temperature). In the case of $MgB_2$, this intrinsic material loss is further complicated by the presence of two superconducting gaps. Although decoupling intrinsic material losses from other external frequency-dependent losses (such as antenna loss) remains challenging, previous studies [28-30,32,33,44,56,68] on the noise temperature of $MgB_2$ HEBs above 1 THz suggest such frequency-dependent material losses may be present. Because of its intrinsic dependence on the material, it is not clear if $\beta$ can be optimized easily, and enhancing this parameter will likely require more extensive investigation of mixer performance across different operating frequencies and different superconducting materials.

Despite uncertainties about how much we can improve on $\alpha_0$ and $\beta$, it is instructive to estimate the theoretical sensitivity limit under near-ideal coupling conditions ($\alpha_0 = \beta = 1$). In this hypothetical scenario, the improved coupling efficiency reduces the required incident LO power to $P_{inc} = 1\mu W$ for the model to achieve the same level of absorbed LO power $P_{LO} \approx 1\ \mu W$ near the optimal bias points > 3 mV. Since the IVC of the HEB is assumed unchanged, the operating conditions in terms of the power ratio $P_{LO}/P_{DC} \approx 2.5$ and the self-heating parameter $C \approx 0.65$ remain the same. Under these conditions, the receiver performance is predicted to improve significantly (Fig. 7(d)): the conversion gain increases to –11 dB (solid black curve) and the receiver noise temperature decreases to 400 K (solid red curve) near the operating point at 2.5 mV. While this is an approximately 8-fold improvement over the current device, the reduced noise temperature is still roughly four times the theoretical limit ($T_M^{CL} + T_M^{QL} \approx 91$ K at 2.5 THz). This gap is in part due to the significant contribution of the Johnson noise $\approx$ 200 K, which only vanishes in the ideal operating conditions $P_{LO} \gg P_{DC}$ and $C \approx 1$.

These results suggest that while optimizing optical coupling is essential, further suppression of Johnson noise is likely necessary for HEB devices to approach the theoretical noise limit. Whereas the complete suppression of the Johnson noise is predicted in a regime where $P_{LO} \gg P_{DC}$ and $C \approx 1$, experimental realization for such a condition can be challenging. One feasible approach is to realize the large $C$ values ($C > 1$) accessible in the negative differential resistance (NDR) region in the IVCs of HEB devices (see the blue curve in Fig. 1(a) for such an IVC and that in Fig.1(b) for the corresponding $C$ values). Under this operating condition, a significant $C$ value is expected to enhance the conversion gain, thereby reducing the Johnson noise contribution. Although these model results to date are strictly derived for uniform HEB devices, we speculate that Johnson noise may also be a significant source of noise in non-uniform HEB devices under similar operating conditions. If this hypothesis holds, operating HEBs in the NDR region may further reduce the noise temperature of NbN HEB devices. Operating NbN HEBs in the NDR regime has been challenging due to instability and MHz oscillations but stabilizing them with custom bias circuits could be feasible. This may offer a relatively straightforward experimental path to approach the theoretical sensitivity limit in HEB mixers [69].

## V. CONCLUSION

In conclusion, we have conducted an experimental study of THz HEB mixers based on sputtered MgB$_2$ thin films and performed a comprehensive model analysis using a simple lumped-element model for HEB mixers. The experimental bias dependence of the conversion gain, noise temperature, and coupled LO power agreed very well with the model analysis in the framework of bias-dependent THz absorption. Our noise analysis indicates that the sensitivity of current HEB devices is limited by large optical losses, with both TEF and Johnson noise contributing to the overall noise temperature. Although improving the RF coupling factor is expected to substantially reduce these contributions, the model results indicate that enhancing the coupling factor alone is insufficient to approach the theoretical sensitivity limit, owing to the residual dominance of Johnson noise. Therefore, achieving lower noise temperatures closer to the quantum limit requires further suppression of Johnson noise, in addition to relatively straightforward RF coupling optimization. One potential route is to explore alternative operating conditions in which the Johnson noise can be significantly reduced by either a high $P_{LO}/P_{DC}$ ratio or a large value of the self-heating parameter $C$. While the former condition is more complex to satisfy experimentally, achieving stable HEB operation in the negative differential resistance region (NDR) with $C \gg 1$ [69,70] is a promising pathway for further optimization of the noise performance of HEB devices.

## ACKNOWLEDGEMENT

This research was conducted at the Jet Propulsion Laboratory, California Institute of Technology, under contract with the National Aeronautics and Space Administration. Changyun Yoo's research was supported by an appointment to the NASA Postdoctoral Program, which Oak Ridge Associated Universities administers under contract with NASA. The authors are grateful to J. Kawamura for providing an FMC source and to C. Curwen for assistance with the THz setup.

# APPENDICES

## APPENDIX A: EFFECTIVE THERMAL CONDUCTANCE FOR SPUTTERED MgB$_2$ THIN FILMS

To determine the effective thermal conductance, we performed isothermal line analysis on the IVCs of one of the fabricated HEB devices (BL29). We assumed that the electron temperature remains constant along the line where the ohmic resistance is constant [40], thereby ignoring the non-thermal dependence $R(I)$ observed in [39]. The IVCs were measured at temperatures ranging from 4.2 to 23.6 K without THz power. The isothermal lines with constant resistances $R_{const}$ = 12 to 30 Ω were drawn on top of the IVCs (Fig. 8(a)). Along these isothermal lines, the Joule power $P_{DC}$ dissipated in the HEB bridge was extracted as a function of the bath temperature $T$ at a fixed $T_e$.

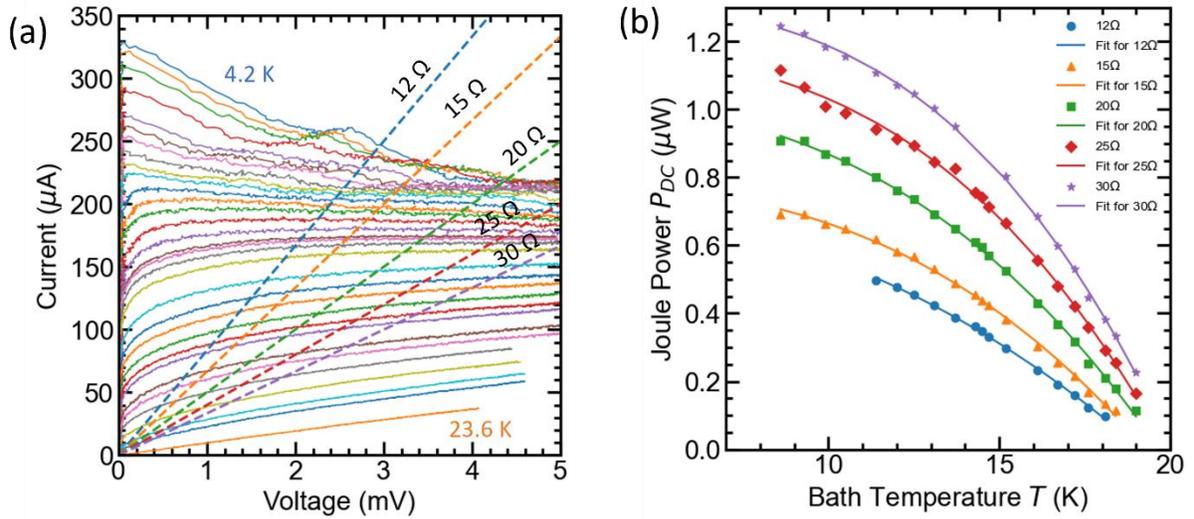

FIG. 8. (a) IVCs of BL29 measured between 4.2 and 23.6 K, along with constant resistance lines with resistances $R_{const}$ = 12-30 Ω. (b) Dissipated Joule power as a function of the bath temperature along each constant resistance line. The symbols show the Joule power values, and the solid lines show the fit using Eq. (A1).

The extracted dissipated Joule power as a function of $T$ each constant resistance line is plotted in Fig. 8(b). For $R_{const}$ = 12 Ω, the data points for $T < 11$ K were disregarded due to the hysteresis of the IVCs at the cross points. For low-temperature hot-electron devices, the following heat balance equation typically applies:

$$P_{DC} + P_{LO} = \Sigma V(T_e^n - T^n), \tag{A1}$$

where $P_{LO}$ is the absorbed LO power (set to 0 for this analysis), $\Sigma$ and $n$ are the material-dependent coefficients, and $V$ is the device volume. Figure 8(b) shows that the extracted dependence $P_{DC}(T_e)$ can be fitted well with Eq. (A1), suggesting $n$ = 2.7-3.0 depending on $R$ (Table II). The effective thermal conductance is then found as $G = dP_{DC}/dT_e = n\Sigma V T_e^{n-1}$.

Table II shows a systematic increase in the $G$ value with increasing resistance $R$, a trend widely observed in previous studies, dating back to early work [39]. This increase in $G$ with $R$ is likely attributed to the expansion of the effective volume $V'$ participating in thermal processes, which grows with the bias (with the device resistance). The underlying mechanism is as follows: the presence of magnetic flux induces a uniform resistive state, and as the bias increases, the volume occupied by the resistive cores of the magnetic vortices expands, leading to a larger $V'$. At higher resistance, $V'$ approaches the total microbridge volume $V$ in Eq. (A1), resulting in higher extracted $G$ values. This interpretation is further supported by the proportional relationship between $G$ and $R$ in Table I; as $R$ increases from 12 Ω to 30 Ω by a factor of 2.5, $G$ value increases from 81.3 nW/K to 208 nW/K by a similar factor ($\approx$ 2.6).

TABLE II. Effective thermal conductance for $MgB_2$ Thin Films Sputtered on $SiN_x$/Si Substrate

| $R_{const}$ (Ω) | $T_e$ (K) | $G$ (nW/K) | $n$ |
|---|---|---|---|
| 12 | 19.1 ± 0.1 | 81.3 | 2.7 ± 0.2 |
| 15 | 19.4 ± 0.1 | 116 | 2.8 ± 0.1 |
| 20 | 19.7 ± 0.1 | 114 | 2.8 ± 0.1 |
| 25 | 20.0 ± 0.1 | 155 | 2.9 ± 0.1 |
| 30 | 20.2 ± 0.2 | 208 | 3.0 ± 0.1 |

**APPENDIX B: BIAS DEPENDENCE OF THE MIXER TIME CONSTANT $\tau^*$**

The time constant was found using a well-known method that measures the modulation frequency response (see, e.g., [71]) of the HEB device to THz radiation. The modulation was created by mixing a quantum cascade laser (QCL) and a tunable FMC source. The power of the FMC source absorbed by the device was kept constant by monitoring the low-frequency (~20 Hz) direct detector response of the HEB devices to optically chopped radiation from the FMC source. Also, a correction was made due to the frequency dependence of the S21 parameter of the IF line. The resulting modulation spectrum $P_{IF}(\omega)$ could be fitted with a Lorentzian function $P_{IF}(\omega) \sim [1 + (\omega\tau^*)^2]^{-1}$; hence, a time constant $\tau^*$ was derived (Fig. 9).

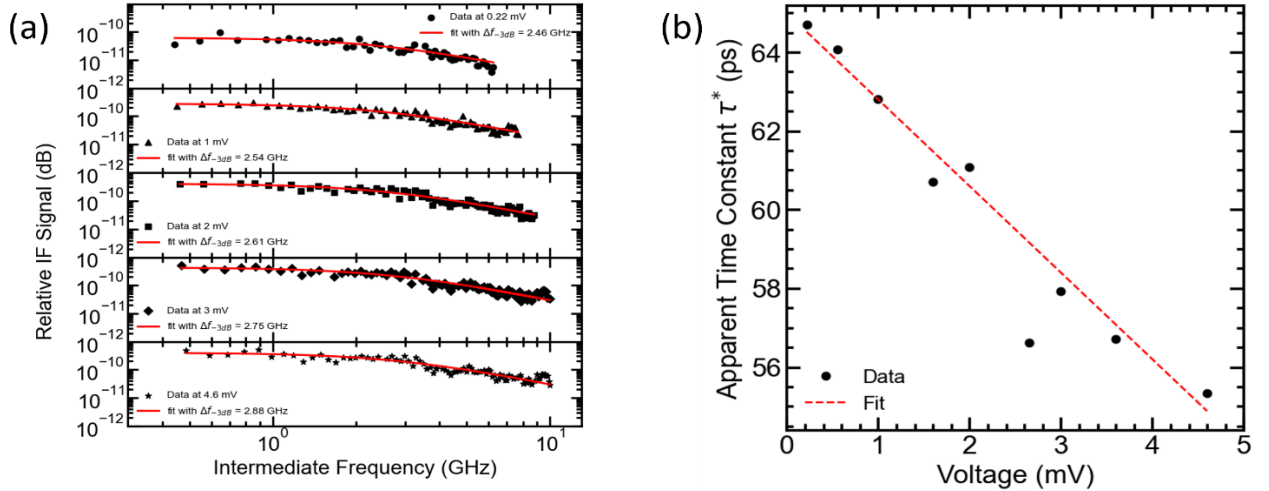

FIG. 9. (a) IF frequency dependence of the mixing over different voltage biases at 5 K. (b) Measured bias dependence of $\tau^*$ (solid circles) along with a linear fit to the data (red dashed line).

Diffusion plays a negligible role, and the time constant is given by $\tau \approx \tau_{eph} + c_e/c_p \cdot \tau_{esc}$. For a lumped-element bolometer with a sufficiently long device length $L \gg \sqrt{D\tau_{eph}}$, where $\tau_{eph}$ is the electron-phonon interaction time, the dominant cooling mechanism for hot electrons is phonon cooling.